\documentclass[review]{elsarticle}
\usepackage{graphicx}
\usepackage{float}
\usepackage{amsmath}
\usepackage{amssymb}
\usepackage{booktabs}
\usepackage{xcolor}
\usepackage{soul}
\usepackage{multirow}
\usepackage{array}
\usepackage{url}
\journal{International Journal of Plasticity}
\biboptions{sort&compress}

\begin{document}

\begin{frontmatter}

\title{Hydrogen-stabilized multimodal high-index twin network in iron}

\author[inst1]{Mehrab Lotfpour}
\author[inst1]{Haoran Cui}
\author[inst1]{Yan Wang}
\author[inst2]{Eduardo Vitral}
\author[inst1]{Lei Cao\corref{cor1}}
\cortext[cor1]{Corresponding author}
\ead{leicao@unr.edu}

\affiliation[inst1]{organization={Department of Mechanical Engineering, University of Nevada, Reno},
            city={Reno},
            postcode={89557},
            state={NV},
            country={USA}}

\affiliation[inst2]{organization={Department of Mechanical Engineering, Rose-Hulman Institute of Technology},
            addressline={5500 Wabash Ave},
            city={Terre Haute},
            postcode={47803},
            state={IN},
            country={USA}}

\begin{abstract}

Hydrogen significantly affects plasticity in bcc iron, but its role is commonly attributed to dislocation-mediated mechanisms, leaving the influence of hydrogen on phase transformation and deformation twinning poorly understood. We use a density-functional-theory-trained deep-neural-network interatomic potential and large-scale molecular dynamics to study pure bcc Fe and Fe containing 10\% hydrogen. It is found that hydrogen lowers the yield stress but, more importantly, increases the persistence of $\{112\}$ twin variants and suppresses detwinning. This effect stabilizes an interconnected $\{332\}$--$\{10\,9\,3\}$ multimodal twin network and produces pronounced post-yield hardening. Higher temperature promotes the initial transformation but weakens the persistence of the high-index multimodal  twin network. Moreover, twinning follows distinct loading-dependent pathways: compression generates $\{332\}$ and $\{10\,9\,3\}$ boundaries through co-zone and non-co-zone twin--twin interactions, whereas tension produces $\{7\,4\,1\}$ boundaries through non-co-zone interactions. These results establish a pathway-based picture in which the intermediate phase determines the accessible twin modes and variant crystallography, while hydrogen and temperature control their kinetic survival and the emergence of high-index twin networks.
\end{abstract}

\end{frontmatter}

\section{Introduction}

Iron (Fe) and Fe-based alloys remain indispensable structural materials for transportation, power generation, manufacturing, and energy infrastructure, while the expansion of hydrogen production, transport, and storage is placing these materials in increasingly hydrogen-rich environments. Their mechanical response is governed by a competition among dislocation slip, deformation twinning, stress-induced phase transformation, and the motion and interaction of the interfaces created by these processes. Deformation twinning is particularly important when ordinary slip is kinetically or geometrically constrained because it can rapidly accommodate strain, reorient the lattice, subdivide grains, and obstruct subsequent dislocation transmission and strain hardening. In body-centered-cubic (bcc) metals, however, twinning is unusually sensitive to loading sense, temperature, stress state, strain rate, and lattice instability, and its atomistic mechanisms remain less settled than in many close-packed metals~\cite{ChristianMahajan1995,LiZhangWang2023}. This sensitivity becomes especially important in ferritic Fe exposed to hydrogen, where small changes in the active plasticity pathway can substantially change the competition between hardening, localization, and fracture~\cite{RobertsonEtAl2015,MartinEtAl2020}.

The conventional description of bcc deformation twinning centers on the $\{112\}\langle111\rangle$ family~\cite{ChristianMahajan1995,LiZhangWang2023}. Yet a growing body of work shows that bcc and metastable-bcc materials can develop much richer twin structures, including $\{332\}$ and other high-index twin boundaries. For example, $\{332\}\langle113\rangle$ twins in metastable $\beta$-titanium (Ti) have been connected to stress-induced martensitic transformation and reverse transformation rather than to direct shearing of the parent bcc lattice~\cite{CastanyEtAl2016}, while high-index twins such as $\{10\,9\,3\}\langle331\rangle$~\cite{antonov2020highindex,gao2025unique} and $\{5\,8\,1\}\langle513\rangle$~\cite{zhang2019strong} have been observed experimentally in metastable $\beta$-Ti alloys. Theoretical work based on symmetry breaking has further shown that fcc-, hcp-, or orthorhombic-like intermediate states can generate distinct bcc twinning paths and naturally produce nonclassical high-index modes~\cite{gao2020twinningpath}. More recently, direct observations of transformation-induced twin boundaries in bcc metals have reinforced the idea that a transient structural state can be an integral part of twin formation rather than a passive by-product~\cite{li2023transitional}. These findings motivate a broader view: the final twin structure and misorientation do not necessarily identify a unique atomic pathway, and high-index boundaries may encode the history of an intervening phase transformation.

A systematic phase-transformation-mediated picture of twinning has emerged from a series of studies by Cao and co-workers. In Mg, $\{10\bar{1}2\}$ and $\{10\bar{1}1\}$ twin nucleation was shown to proceed through an hcp$\rightarrow$bcc$\rightarrow$hcp twin transformation, in which an evanescent metastable phase provides the crystallographic bridge between parent and twin~\cite{OmbogoEtAl2020, ombogo2025nucleation}. In Ti, $\{11\bar{2}2\}$ contraction twinning was found to occur through an $\alpha\rightarrow\omega\rightarrow\alpha$ pathway~\cite{zahiri2021formation}, and stress-induced $\omega\rightarrow\alpha$ transformation was shown to generate several distinct transformation-twin families~\cite{ZahiriEtAl2021Omega}. The same framework also revealed that interactions between non-co-zone $\{10\bar{1}2\}$ variants in Mg generated $\{11\bar{2}2\}$ twin--twin boundaries~\cite{zahiri2022formation}. Extending this concept to deformation twinning in bcc materials, Cao and co-workers demonstrated that nominally similar $\{112\}$ twins can arise through hcp or fcc intermediate states, with different shear magnitudes, shuffle contents, temperature dependence, and subsequent twin--twin interactions~\cite{zahiri2024anisotropy}. These results led to a more general framework in which the loading direction and temperature govern the formation of the intermediate phase and the kinetic pathway to the final twin~\cite{zahiri2023twinning,cao2026deformation}.

Iron provides a stringent test of this picture because bcc, fcc, and hcp configurations can become mechanically competitive. Atomistic simulations have shown that bcc Fe can transform into either hcp- or fcc-intermediate structures depending on the temperature~\cite{ShaoEtAl2018}. Hydrogen introduces a second, chemically driven perturbation to this transformation landscape. Hydrogen embrittlement in ferritic steels is not associated with a single universal mechanism; hydrogen can modify dislocation mobility and interactions, segregate to defects and interfaces, alter vacancy stability and cohesion, and change crack-tip plasticity, with the dominant response depending strongly on stress state, temperature, microstructure, and hydrogen concentration~\cite{RobertsonEtAl2015,MartinEtAl2020,KumarEtAl2023}. Recent molecular dynamics simulations showed that hydrogen pinning of edge dislocations can increase the local shear stress sufficiently to initiate microtwinning~\cite{Matsumoto2024}, demonstrating that hydrogen can promote twinning indirectly through its effect on defect kinetics. Nevertheless, direct investigations into the interplay between hydrogen and twinning in bcc Fe remain comparatively limited.

Resolving this problem requires an atomistic description that is accurate across several Fe phases and Fe-H environments while remaining efficient enough to capture collective deformation processes. Density-functional theory can describe competing Fe phases and Fe-H bonding with first-principles fidelity, but the system sizes and time scales required for collective phase transformation, twin growth, twin--twin impingement, and detwinning are far beyond routine direct DFT dynamics. Conventional empirical potentials permit large-scale molecular dynamics but may lose transferability when a single simulation traverses highly strained bcc states, transient hcp and fcc environments, twin boundaries, disordered configurations, and interstitial H. Deep-neural-network interatomic potentials provide a route to bridge these scales by learning a high-dimensional potential-energy surface from first-principles reference data while retaining near-classical molecular-dynamics efficiency~\cite{ZhangEtAl2018DPMD,wang2018deepmd}. Their suitability for Fe-H has been demonstrated by neural-network and other machine-learning potentials capable of treating hydrogen interactions with bulk defects, dislocations, grain boundaries, segregation sites, and fracture environments with substantially improved transferability over conventional analytic potentials~\cite{MengEtAl2021,ItoEtAl2026}. A potential intended for transformation-mediated twinning must additionally represent the competing bcc, hcp, and fcc environments and the strongly distorted configurations that connect them.

In this work, we use a first-principles-trained deep-neural-network potential and large-scale molecular dynamics simulations to resolve how hydrogen, temperature, and loading sense reorganize twinning pathways in bcc Fe. Pure Fe and FeH$_{0.1}$ are subjected to $[100]$ compression and tension, with the temperature varied between 300 and 600~K. The paper is organized as follows. Section~\ref{sec:method} describes the methodologies employed in this study. Section~\ref{sec:results} presents the results of extensive MD simulations and integrates the simulation observations with theoretical calculations. The dependence of the deformation and twinning mechanisms on loading direction is further discussed in Section~\ref{sec:discussion}. Finally, Section~\ref{sec:conclusions} summarizes the main findings and provides concluding remarks.

\section{Methodology}\label{sec:method}

\subsection{Ab initio molecular dynamics}

The reference configurations for the deep-neural-network (DNN) potential were generated by ab initio molecular dynamics (AIMD) using the Vienna~\textit{Ab initio} Simulation Package (VASP)~\cite{kresse1996vasp}. Electron--ion interactions were described by the projector-augmented-wave method~\cite{blochl1994paw}, and exchange--correlation effects were treated within the Perdew--Burke--Ernzerhof generalized-gradient approximation~\cite{perdew1996pbe}. Spin polarization was included for Fe-containing systems. The plane-wave cutoff energy was 600~eV, the electronic convergence criterion was $5.0\times10^{-7}$~eV, and Gaussian smearing with a width of 0.05~eV was applied. Because the AIMD supercells were comparatively large, Brillouin-zone sampling was restricted to the $\Gamma$ point. Dispersion interactions were included using the DFT-D3 correction with Becke--Johnson damping~\cite{grimme2011d3}.

The database was designed to span the broad Fe--O--H chemical and structural space. Specifically, it includes bcc Fe, fcc Fe, hcp Fe, Fe-H, Fe--O, bulk Fe--water configurations, and $\mathrm{Fe_3O_4}$. AIMD trajectories sampled temperatures from 10 to 2500~K and pressures from $-5$ to 150~GPa, thereby covering equilibrium crystals, thermally disordered structures, strong compression, and expanded or tensile states. A Langevin thermostat and a 1.0~fs integration step were used. Lattice vectors, atomic coordinates, total energies, and atomic forces were extracted from the trajectories for potential training and validation.

\subsection{DNN potential training and validation}

The potential was trained using the DeePMD-kit framework~\cite{wang2018deepmd}. The local-environment cutoff radius was 6.0~\AA, with a 0.5~\AA{} smooth-transition region. The embedding network contained hidden layers of 32, 64, and 128 neurons, and the fitting network contained three hidden layers of 128 neurons each. Hyperbolic-tangent activation functions were used throughout. The learning rate decayed exponentially from $1.0\times10^{-3}$ to $3.51\times10^{-8}$ over 16 million training steps. During training, the prefactor of the energy loss increased from 0.02 to 1, whereas the force prefactor decreased from 1000 to 1. This schedule prioritizes force accuracy during the early exploration of the potential-energy surface and progressively balances energies and forces as convergence is approached.

\begin{figure}[H]
    \centering
    \includegraphics[width=1\textwidth]{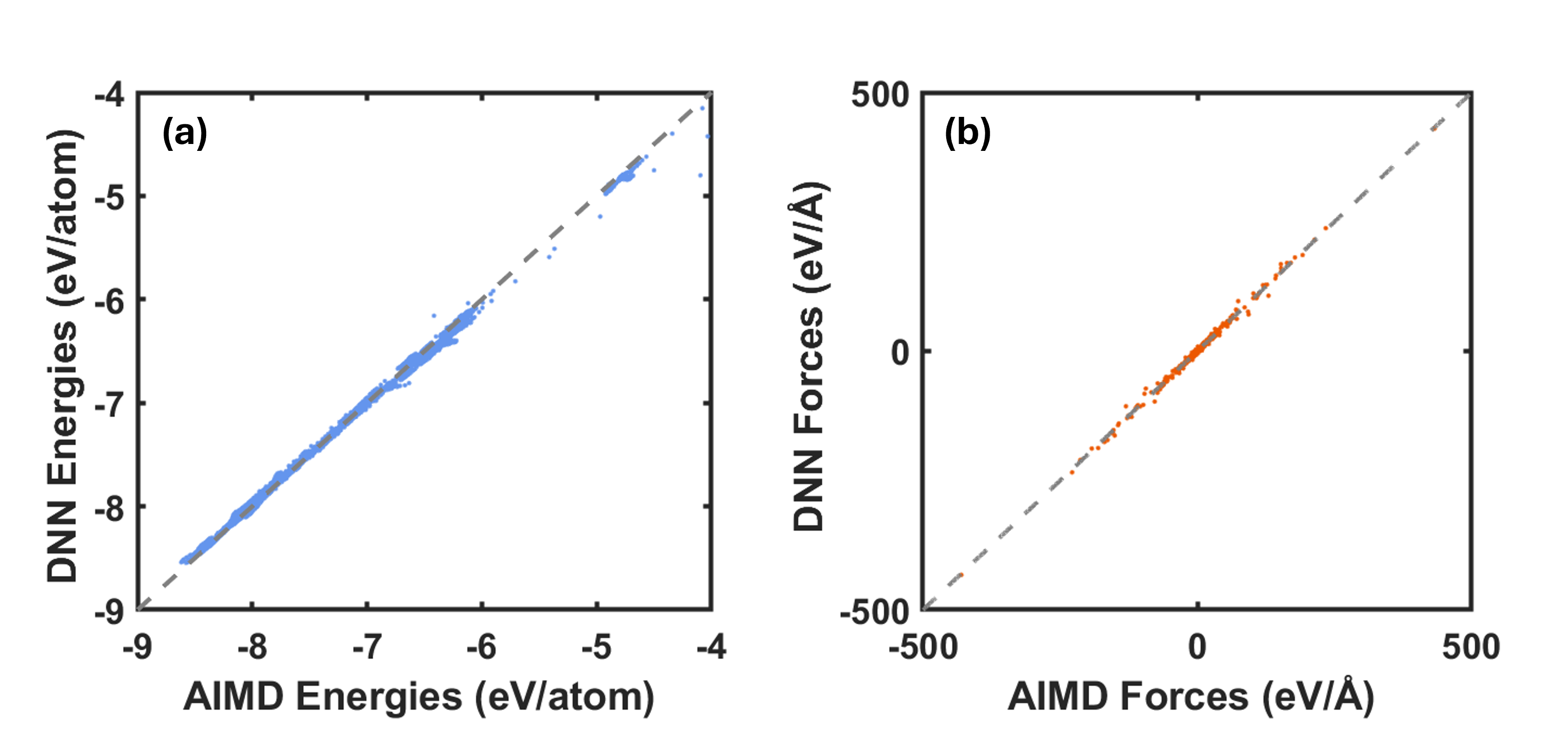}
    \caption{Validation of the DNN interatomic potential against AIMD reference data. Parity plots compare (a) energies and (b) atomic-force components. Dashed lines denote ideal agreement. }
    \label{fig:AIMD_DNN}
\end{figure}

Figure~\ref{fig:AIMD_DNN} compares DNN predictions with AIMD reference energies and forces. The data remain concentrated around the parity lines over the sampled range, indicating that the potential reproduces the first-principles database without systematic bias. This agreement is particularly important here because the deformation trajectories traverse strongly strained bcc states, transient close-packed configurations, and Fe-H environments that are difficult to represent simultaneously with conventional analytic potentials.

\subsection{DNN-MD deformation simulations and structural analysis}

Single-crystal bcc Fe models were oriented as x=$[100]$, y=$[010]$, and z=$[001]$, so that the loading axis was parallel to $[100]$. Two compositions were examined: pure Fe and Fe containing a nominal 9.09~at.\% interstitial H, in which the number of Fe and H atoms have a ratio of 10:1 and thus will be referred to as $\mathrm{FeH}_{0.1}$. Simulation cells were constructed as $48\times48\times48$ periodic replications of the bcc unit cell (lattice constant $a = 2.8047$~\AA), giving 243{,}293 atoms in total (221{,}184 Fe; for the hydrogen-containing composition, an additional 22{,}109 interstitial H atoms placed at randomly selected tetrahedral-type sites with an independent 10\% probability per Fe lattice site, yielding a nominal H:Fe ratio of 1:10), with periodic boundary conditions applied in all three directions.

The MD simulations were then performed in the LAMMPS package~\cite{plimpton1995fast}. Each configuration was first energy-minimized with the simulation cell allowed to relax anisotropically to zero pressure, then equilibrated for 500~ps in the NPT ensemble at the target temperature (300 and 600~K) using a Nos\'e--Hoover thermostat~\cite{nose} and an anisotropic Nos\'e--Hoover barostat (target pressure 0~bar)~\cite{baro}. Uniaxial compression or tension was then applied at a strain rate of $10^{9}$~s$^{-1}$ at 300 and 600~K for 500~ps (1~fs timestep) along the loading axis. The axial virial stress was recorded to construct the stress--strain curves.

The crystal structures and orientation variants were analyzed in OVITO~\cite{ovito}. Atoms were classified as bcc, fcc, hcp, and~\textit{Other} using common neighbor analysis (CNA)~\cite{cna1,cna2}. For variant identification, interstitial H atoms were removed, and Polyhedral Template Matching (PTM)~\cite{larsen2016robust} was used to assign per-atom local orientations, which were then grouped into discrete grains using the Grain Segmentation modifier~\cite{ovito} with an automatic merge threshold and a minimum grain size of 50 atoms.

\section{Results}

\label{sec:results}

\begin{figure}[H]
    \centering
    \includegraphics[width=\textwidth]{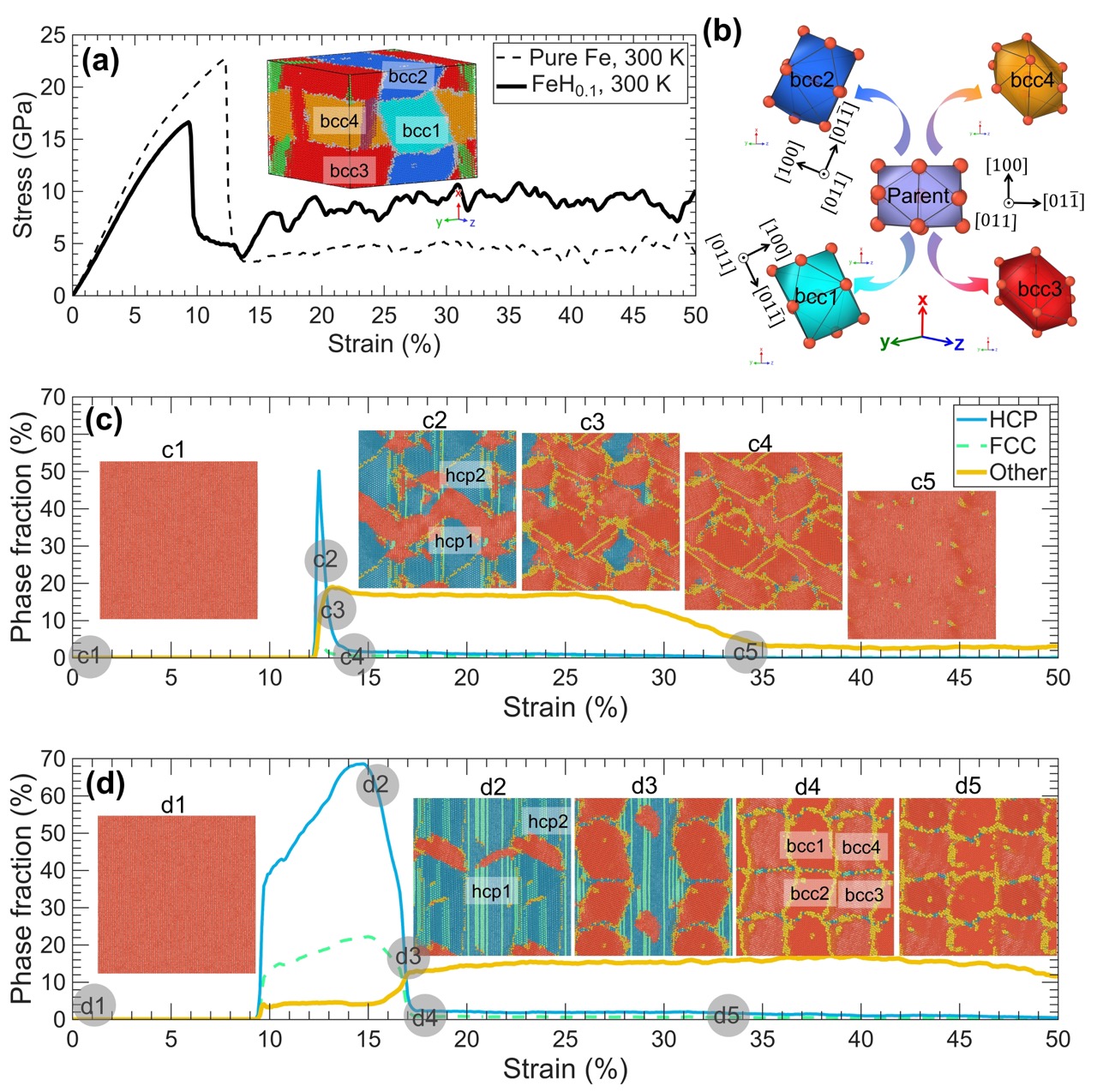}
    \caption{
    Deformation and phase evolution of pure Fe and $\mathrm{FeH}_{0.1}$ under $[100]$ compression at 300~K and a strain rate of $10^{9}$~s$^{-1}$.
    (a) Stress--strain curves of pure Fe and $\mathrm{FeH}_{0.1}$, showing that hydrogen reduces the stress associated with the first structural instability but produces a substantially higher post-instability flow stress.
    (b) Crystallographic correspondence between the parent bcc lattice and the four bcc twin variants.
    (c,d) Evolution of the hcp, fcc, and structurally disordered (\textit{Other}) phase fractions in pure Fe and $\mathrm{FeH}_{0.1}$, respectively. The bcc fraction, which constitutes the remainder of the system, is omitted to better resolve the minority phases. The atomistic snapshots show representative microstructures at the strains marked on the phase-fraction curves and are viewed along $[0\bar{1}1]$.
    }
    \label{fig:stress_strain}
\end{figure}

We first examine the deformation and twinning mechanisms of pure Fe and H-containing Fe (denoted as $\mathrm{FeH}_{0.1}$) subjected to uniaxial compression at 300~K. Figure~\ref{fig:stress_strain}a compares the stress--strain responses of pure Fe and $\mathrm{FeH}_{0.1}$. Both systems initially exhibit a steep elastic response followed by an abrupt stress drop. Hydrogen shifts the yielding to a lower stress and strain. After yielding, however, $\mathrm{FeH}_{0.1}$ sustains a markedly higher flow stress than pure Fe over nearly the entire simulated strain range. The notable hydrogen-induced hardening is rooted in the complex multimodal  twin network formed, as shown by the inset over the stress--strain curve.

Figure~\ref{fig:stress_strain}c shows the evolution of the non-bcc phase fractions in pure Fe together with representative microstructures as insets. At a strain of approximately 12.5\%, coincident with the abrupt stress drop in Fig.~\ref{fig:stress_strain}a, nearly one-half of the bcc Fe transforms collectively into metastable hcp domains. Interestingly, the transformed region contains two dominant hcp orientation variants, denoted hcp1 and hcp2. These variants exhibit a $\{10\bar{1}2\}$ twin relationship, characterized by an approximately $86^{\circ}$ misorientation about a common $\langle a\rangle$ axis. Their paired formation is not accidental but follows from the crystallographic degeneracy of the transformation pathways available to the parent bcc lattice, as discussed in our previous work~\cite{zahiri2022role}.

The hcp domains are highly transient. They begin to transform back into bcc at approximately 13.2\% strain (configuration c2 in Fig.~\ref{fig:stress_strain}c) and are almost completely consumed by approximately 14.0\% strain (configuration c4 in Fig.~\ref{fig:stress_strain}c). Notably, each hcp orientation produces more than one bcc product: hcp1 transforms into the bcc1 and bcc2 variants, whereas hcp2 transforms into bcc3 and bcc4. As illustrated in Fig.~\ref{fig:stress_strain}b, each newly formed bcc variant is related to the parent bcc lattice through a $\{112\}$ twin relationship. Similar transformation-mediated twinning mechanisms have been identified in previous studies of Fe~\cite{zahiri2024anisotropy}, Ti~\cite{zahiri2022role}, Mg~\cite{zahiri2021formation, OmbogoEtAl2020}, and Ti--Nb alloys~\cite{zahiri2024anisotropy} and were shown to govern important microstructural phenomena, including twin-variant selection, tension--compression asymmetry, and the temperature dependence of twinning behavior~\cite{zahiri2023twinning}. 

Upon continued compression, the newly formed bcc twin variants initially continue to grow until approximately 29\% strain. At larger strains, however, the twin variants begin to shrink, and pronounced detwinning becomes evident near 33.5\% strain, as illustrated by configuration c5 in Fig.~\ref{fig:stress_strain}c. The simultaneous decrease in the~\textit{Other} fraction and recovery of the bcc fraction are consistent with the detwinning process. Finally, detwinning leaves a predominantly parent-bcc structure in which continued plastic deformation can proceed at a comparatively low flow stress.

\subsection{Effect of hydrogen}

The microstructural response of $\mathrm{FeH}_{0.1}$ differs fundamentally from that of pure Fe. As shown in Fig.~\ref{fig:stress_strain}d, the hcp phase first appears at approximately 9.5\% strain, substantially earlier than in pure Fe, and remains detectable until approximately 17.5\% strain. The corresponding life span of the hcp phase is more than five times that in pure Fe. The hcp fraction also reaches a larger maximum value, demonstrating that hydrogen promotes both the spatial extent and the kinetic persistence of the hcp intermediate phase. It is expected that interstitial hydrogen introduces heterogeneous local lattice distortions and alters the relative stability of the mechanically competing bcc and hcp configurations. These local perturbations provide favorable sites for transformation embryos and reduce the barrier for the collective structural rearrangement. The earlier onset of the hcp phase in Fig.~\ref{fig:stress_strain}d is consequently consistent with the lower yield stress of $\mathrm{FeH}_{0.1}$ in Fig.~\ref{fig:stress_strain}a.

The most striking effect of hydrogen, however, is exhibited in the flow stress. In pure Fe, the $\{112\}$ twin networks eventually shrink and disappear through detwinning. In $\mathrm{FeH}_{0.1}$, all four $\{112\}$ twin variants remain stable and undergo extensive growth. As shown by configurations d2--d4 in Fig.~\ref{fig:stress_strain}d, the variants expand until neighboring twin domains impinge and form extended twin--twin interfaces throughout the simulation cell. Twin variant pairs derived from the same hcp intermediate, namely bcc1/bcc2 (or bcc3/bcc4), form $\{332\}$-type twin boundaries. In contrast, twin variants derived from the two different hcp intermediates, namely bcc1/bcc4 (or bcc2/bcc3), produce $\{10\,9\,3\}$ high-index twin boundaries.

The resulting microstructure forms a three-dimensional, mutually constraining network of $\{332\}$ and $\{10\,9\,3\}$ twin boundaries. The approximately horizontal interfaces in Fig.~\ref{fig:stress_strain}d are predominantly associated with the $\{332\}$ relationships, whereas the approximately vertical interfaces correspond to the high-index $\{10\,9\,3\}$ boundaries. Because the two boundary families intersect repeatedly, their migration cannot occur independently. Motion of one boundary requires coordinated rearrangement of the adjoining boundaries and their junctions, producing strong geometrical pinning. The evolution from configuration d4 to d5 in Fig.~\ref{fig:stress_strain}d illustrates this constraint. Rather than translating freely, the nominally horizontal $\{332\}$ boundaries bow between intersections, while the adjoining high-index boundaries straighten and resist lateral displacement. The network therefore redistributes local stresses through boundary bowing and junction motion instead of undergoing rapid, collective detwinning as in the case without hydrogen. Meanwhile, interstitial hydrogen could further reduce boundary mobility by local distortion at the twin boundaries and junctions. The combined effects of mutual geometrical pinning and interstitial hydrogen suppress the detwinning observed in pure Fe. Consequently, the multimodal  twin network remains clearly identifiable at approximately 17\% strain and persists, although increasingly distorted, to at least 44.5\% strain. Continued deformation must therefore proceed in the presence of a stable population of twin boundaries and twin--twin junctions. These interfaces obstruct dislocation motion, constrain the migration of neighboring twins, and require additional stress for further microstructural rearrangement. Their persistence explains why $\mathrm{FeH}_{0.1}$ maintains a substantially higher flow stress than pure Fe. Similar high strain-hardening has been attributed to the dense network of five-fold twins in fcc Cu~\cite{zahiri2019}.

Overall, hydrogen has a dual and strongly coupled influence on the compressive response of Fe. It first promotes the bcc-to-hcp phase transformation, thereby lowering the stress required for transformation-mediated twin nucleation. It then stabilizes the resulting $\{112\}$ twin variants and enables the formation of an interconnected $\{332\}$--$\{10\,9\,3\}$ twin-boundary network. By suppressing twin-boundary migration and detwinning, this multimodal  twin network produces the pronounced hardening observed in $\mathrm{FeH}_{0.1}$.

\subsection{Twinning calculation\label{sec:highindex}}
In this section, we examine in greater detail the crystallography of these complex twin networks. In particular, the present work provides, to the best of our knowledge, the first atomistic observation of $\{10\,9\,3\}$ high-index twin boundaries. 

\begin{figure}[H]
    \centering
\includegraphics[width=0.6\textwidth]{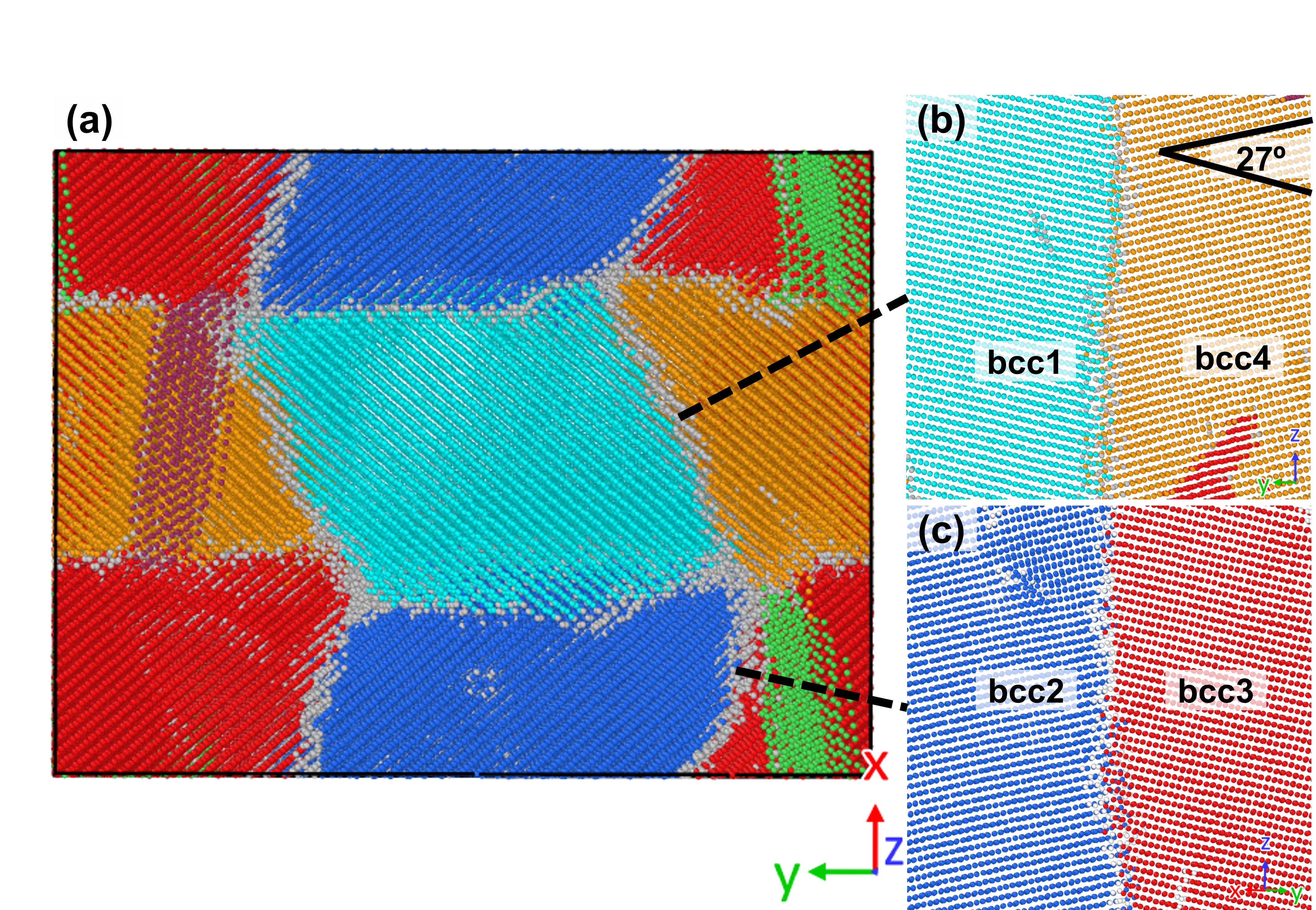}
    \caption{Crystallographic analysis of H-containing Fe during compression. (a) The microstructure showing the interconnected multimodal  twin network. (b) bcc1 and bcc4 share a $[1\bar{1}0]$ zone axis and are misoriented by approximately $27^{\circ}$. (c) bcc2 and bcc3 share a $[110]$ zone axis with a similar misorientation.}
    \label{fig:high_index_comp2}
\end{figure}

\begin{figure}[H]
    \centering
\includegraphics[width=0.8\textwidth]{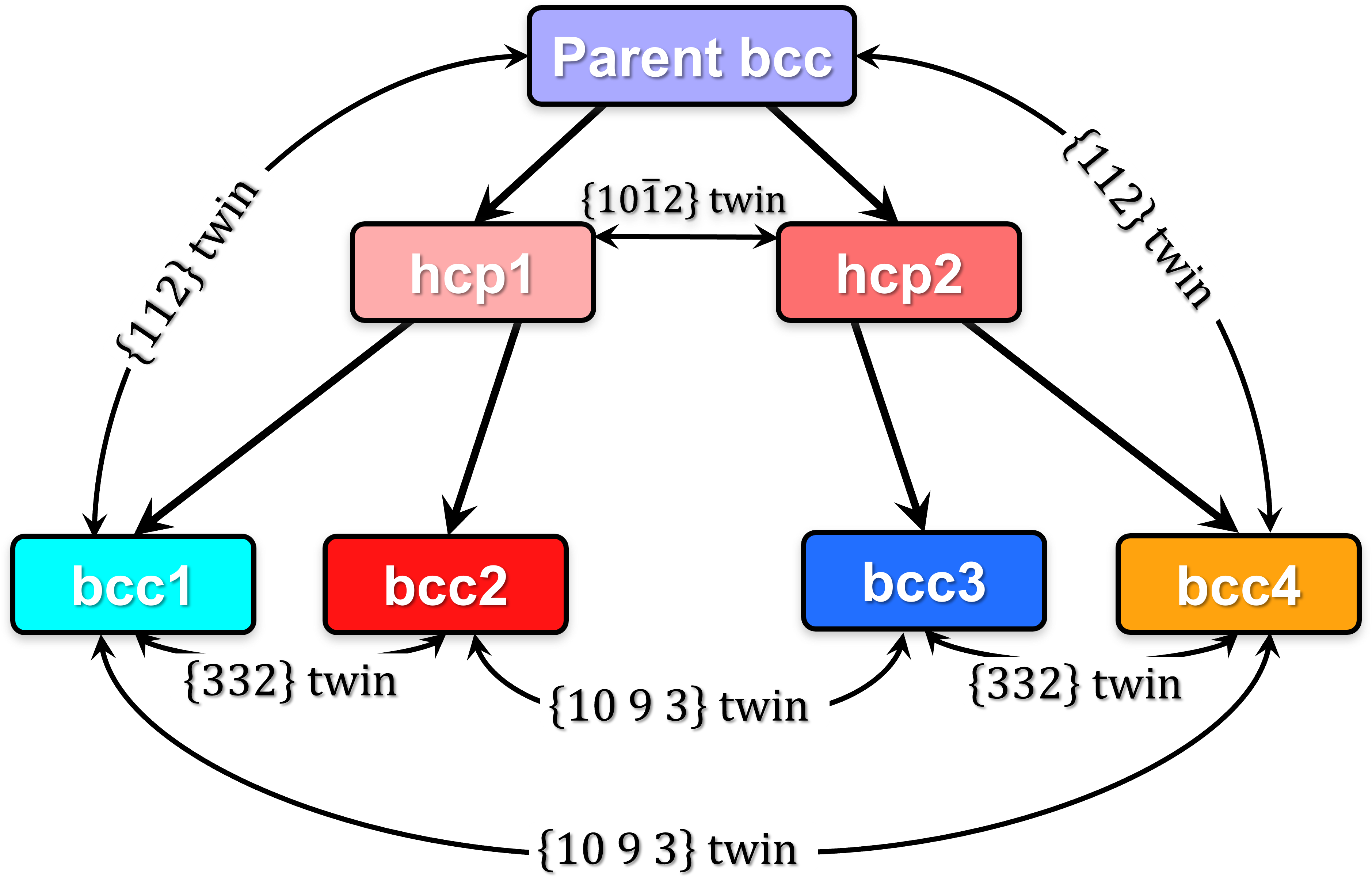}
    \caption{Crystallographic relationships among the parent bcc phase, the two intermediate hcp phases (hcp1 and hcp2), and the four $\{112\}$ bcc twin variants (bcc1, bcc2, bcc3, and bcc4).}
    \label{fig:tree-hcp}
\end{figure}

Figure~\ref{fig:high_index_comp2} presents the grain-segmentation analysis of the four bcc variants generated in $\mathrm{FeH}_{0.1}$ at 300~K. As shown in Fig.~\ref{fig:tree-hcp}, bcc1 and bcc2 originate from intermediate phase hcp1, while bcc3 and bcc4 originate from intermediate phase hcp2. Therefore, bcc1/bcc2 are co-zone $\{112\}$ twins and the corresponding twin--twin interaction forms the horizontal $\{332\}$ twin boundaries. This twin--twin relationship follows the multiplicity of crystallographically compatible bcc products of the reverse hcp$\rightarrow$bcc transformation. 

In contrast, the cross-pairs bcc1/bcc4 originate from different hcp intermediates and are non-co-zone $\{112\}$ twins. In other words, since hcp1/hcp2 are themselves $\{10\bar{1}2\}$-twin-related, the bcc1 variant generated from hcp1 and the bcc4 generated from hcp2 inherit both the $\{10\bar{1}2\}$ hcp twin and the $\{112\}$ bcc twin correspondences. As shown by Fig.~\ref{fig:high_index_comp2}, both bcc1/bcc4 and bcc2/bcc3 exhibit an angle of approximately $27^{\circ}$ around $\langle110\rangle$ common zone axes. This relation is consistent with the reported $\{10\,9\,3\}\langle\bar{3}31\rangle$ high-index bcc twin mode, whose characteristic misorientation is approximately $26^{\circ}$~\cite{antonov2020highindex,LiZhangWang2023}. Besides the misorientation angle, MD simulations allow us to apply quantitative calculations to confirm the complete twinning elements. 

The correspondence matrix $C_{ij}$ characterizes the twinning deformation in the parent bcc lattice basis relative to the twin bcc lattice basis~\cite{ChristianMahajan1995, cayron2019transformation}. But it should be noted that $C_{ij}$ departs from the actual twinning deformation gradient by a rotation due to the change of basis. Generally, twinning does not involve a significant rigid-body rotation, allowing us to focus on the polar decomposition and the resulting stretch tensor, through $\mathbf{U} = \sqrt{\mathbf{C}^\top\cdot\mathbf{C}}$. We denote $\mathbf{G}_i$ and $\mathbf{g}_i$ as the parent and twin lattice bases, respectively. Then the correspondence matrix reads $\mathbf{C} = \mathbf{g}_i\otimes\mathbf{G}^i$, where $\mathbf{G}^i$ represents the reciprocal parent basis satisfying $\mathbf{G}_i\cdot\mathbf{G}^j = \delta_i^j$. From the MD simulations in Fig.~\ref{fig:high_index_comp2}, we track three orthogonal directions and find one set of orientation relations as follows:

\begin{equation}
\begin{split}
    &[100]_{bcc} \quad\leftrightarrow\quad \frac{1}{3}[\bar{1}\bar{1}20]_{hcp1} \quad\leftrightarrow\quad \frac{1}{2}[1\bar{1}\bar{1}]_{bcc1}
    \\[2mm]
    &[011]_{bcc} \quad\leftrightarrow\quad \phantom{\frac{1}{2}}[1\bar{1}00]_{hcp1} \quad\leftrightarrow\quad \frac{1}{2}[311]_{bcc1}
    \\[2mm]
    &[0\bar{1}1]_{bcc} \quad\leftrightarrow\quad \phantom{\frac{1}{2}}[0001]_{hcp1} \quad\leftrightarrow\quad \phantom{\frac{1}{2}}[0\bar{1}1]_{bcc1}\,.
\end{split}
\end{equation}

Starting from the standard orthonormal basis $\mathbf{e}_i$ we define the parent basis as $\mathbf{G}_1 = a_0\,\mathbf{e}_1$, $\mathbf{G}_2 = a_0\,(\mathbf{e}_2+\mathbf{e}_3)$ and $\mathbf{G}_3 = a_0\,(-\mathbf{e}_2+\mathbf{e}_3)$, and the twin basis as $\mathbf{g}_1 = (a_0/2)\,(\mathbf{e}_1-\mathbf{e}_2-\mathbf{e}_3)$, $\mathbf{g}_2 = (a_0/2)\,(3\mathbf{e}_1+\mathbf{e}_2+\mathbf{e}_3)$ and $\mathbf{g}_3 = a_0\,(-\mathbf{e}_2+\mathbf{e}_3)$. The correspondence matrix $C_{ij}$ for the bcc1 twin variant becomes
\begin{equation}
\begin{split}
C_{ij}|_{bcc-hcp1-bcc1} = \left[
      \begin{array}{ccc}
        0.5 \quad & \quad 0.75 \quad & \quad 0.75\\
        -0.5 \quad & \quad 0.75 \quad & \quad -0.25\\
        -0.5 \quad & \quad -0.25 \quad & \quad 0.75\\
      \end{array}
    \right]\,.
\end{split}
\label{eq:dg3}
\end{equation}

For its co-zone twin variant, bcc2 has the correspondence matrix of 

\begin{equation}
\begin{split}
C_{ij}|_{bcc-hcp1-bcc2} = \left[
      \begin{array}{ccc}
       0.5 \quad & \quad -0.75 \quad & \quad -0.75\\
        0.5 \quad & \quad 0.75 \quad & \quad -0.25\\
        0.5 \quad & \quad -0.25 \quad & \quad 0.75\\
      \end{array}
    \right]\,.
\end{split}
\label{eq:dg33}
\end{equation}
For the twin variants bcc3 and bcc4 originating from hcp2, the correspondence matrices read
\begin{equation}
\begin{split}
    C_{ij}|_{bcc-hcp2-bcc3} = \left[
      \begin{array}{ccc}
        0.5 \quad & \quad -0.75 \quad & \quad 0.75\\
        0.5 \quad & \quad 0.75 \quad & \quad 0.25\\
        -0.5 \quad & \quad 0.25 \quad & \quad 0.75\\
      \end{array}
    \right]\,,
    \\[5mm]
    C_{ij}|_{bcc-hcp2-bcc4} = \left[
      \begin{array}{ccc}
        0.5 \quad & \quad 0.75 \quad & \quad -0.75\\
        -0.5 \quad & \quad 0.75 \quad & \quad 0.25\\
        0.5 \quad & \quad 0.25 \quad & \quad 0.75\\
      \end{array}
    \right]\,.    
\end{split}
\label{eq:dg4}
\end{equation}

These correspondence matrices are used to solve the twinning equation following Ball and James~\cite{ball1987fine}, as well as Bhattacharya~\cite{bhattacharya2003microstructure}. Between two twin variants $I$ and $J$, with right stretch tensors $\mathbf{U}_I$ and $\mathbf{U}_J$, the twinning equation takes the form
\begin{equation}
    \mathbf{Q}\cdot\mathbf{U}_I-\mathbf{U}_J = \mathbf{a}\otimes\mathbf{n},
    \label{eq:te}
\end{equation}
for some vector $\mathbf{a}$ and a unit normal $\mathbf{n}$ to the habit plane. The rotation tensor $\mathbf{Q}$ characterizes the misorientation between the two variants.

The twinning direction $\boldsymbol\eta$, habit plane $\mathbf{K}$ and shear magnitude $s$ can be calculated from the elements of~\eqref{eq:te} by
\begin{equation}
    \boldsymbol\eta = \frac{\mathbf{a}}{|\mathbf{a}|}\,, \quad\quad
    \mathbf{K} = \frac{\mathbf{U}^{-1}_J\cdot\mathbf{n}}{|\mathbf{U}^{-1}_J\cdot\mathbf{n}|}\,, \quad\quad
    s = |\mathbf{a}|\,|\mathbf{U}^{-1}_J\cdot\mathbf{n}|\,.
    \label{eq:te2}
\end{equation}

When one of the variants is the reference lattice, say variant $J$, then $\mathbf{U}_J = \mathbf{I}$, which allows us to combine Eqs.~\eqref{eq:te} and \eqref{eq:te2} as follows
\begin{equation}
    \mathbf{Q}\cdot\mathbf{U}_I-\mathbf{I} = s(\boldsymbol\eta\otimes\mathbf{K})\,.
    \label{eq:te3}
\end{equation}
For the horizontal twin--twin boundaries between bcc1/bcc2, the stretch $\mathbf{U}_I$ is calculated from the total correspondence $\mathbf{C}_{bcc-hcp1-bcc1}\cdot\mathbf{C}_{bcc-hcp1-bcc2}^{-1}$. Based on this stretch, from Eq.~\eqref{eq:te3} we obtain
\begin{equation}
\begin{split}
    \mathbf{K}_1 = (\bar{2}\bar{3}\bar{3})\,,\quad
    \boldsymbol\eta_1 = [3\bar{1}\bar{1}]\,,\quad
    s_1 =0.3536\,,
    \\[2mm]
    \mathbf{K}_2 = (\bar{2}11)\,,\quad
    \boldsymbol\eta_2 = [111]\,,\quad
    s_2 =0.3536\,.
    \label{eq7}
\end{split}
\end{equation}
Indeed, the $\{332\}$ twin is characterized by a misorientation of $51.1^{\circ}$ about $\langle1\bar{1}0\rangle$, consistent with the observation in Fig.~\ref{fig:high_index_comp2}.  The calculation for the other co-zone $\{112\}$ twin pair, bcc3/bcc4, also forms the $\{332\}$ twin in the same family. 

We next consider the non-co-zone $\{112\}$ twin variants, bcc1/bcc4. The stretch $\mathbf{U}_I$ is calculated from the total correspondence $\mathbf{C}_{bcc-hcp1-bcc1}\cdot\mathbf{C}_{bcc-hcp2-bcc4}^{-1}$. Based on this stretch, from Eq.~\eqref{eq:te3} we obtain
\begin{equation}
\begin{split}
    \mathbf{K}_1 = (\bar{10}\,\bar{9}\,\bar{3})\,,\quad
    \boldsymbol\eta_1 = [3\bar{3}\bar{1}]\,,\quad
    s_1 =0.3953\,,
    \\[2mm]
    \mathbf{K}_2 = (\bar{2}31)\,,\quad
    \boldsymbol\eta_2 = [531]\,,\quad
    s_2 =0.3953\,.
    \label{eq8a}
\end{split}
\end{equation}
Indeed, the $(\bar{10}\,\bar{9}\,\bar{3})$ twin is characterized by a misorientation of $27^{\circ}$ about $\langle1\bar{1}0\rangle$, consistent with the observation in Fig.~\ref{fig:high_index_comp2}.  The calculation for the other non-co-zone $\{112\}$ twin pair, bcc2/bcc3, also forms the $\{10\, 9\, 3\}$ twin in the same family. 

The analogy with hcp non-co-zone twin--twin interactions should be noted here. In hcp Mg, an $\{11\bar{2}2\}$ twin was recently reported~\cite{cayron2018evidence}. However, detailed MD simulations have demonstrated that this $\{11\bar{2}2\}$ twin is in fact distinct from the $\{11\bar{2}2\}$ deformation twin observed in hcp Ti and Zr. It instead arises from the twin--twin interaction of two non-co-zone $\{10\bar{1}2\}$ twin variants~\cite{zahiri2022formation}. Here, the situation in the bcc system is analogous: co-zone $\{112\}$ twins form $\{332\}$ twins, whereas non-co-zone $\{112\}$ twins form $\{10\,9\,3\}$ high-index twins.

\subsection{Effect of temperature}
\label{sec:compression_600}
The previous section demonstrates that hydrogen plays a key role in enhancing twin stability and suppressing detwinning, thereby promoting the formation of multimodal , high-index twin networks and, consequently, increasing the flow stress. Increasing temperature, on the other hand, might oppose the hydrogen-induced effect through enhanced dislocation slip and reduced hcp phase stability. In this section, we will examine the competition between hydrogen and temperature by increasing the temperature to 600~K.

\begin{figure}[H]
    \centering
    \includegraphics[width=0.95\textwidth]{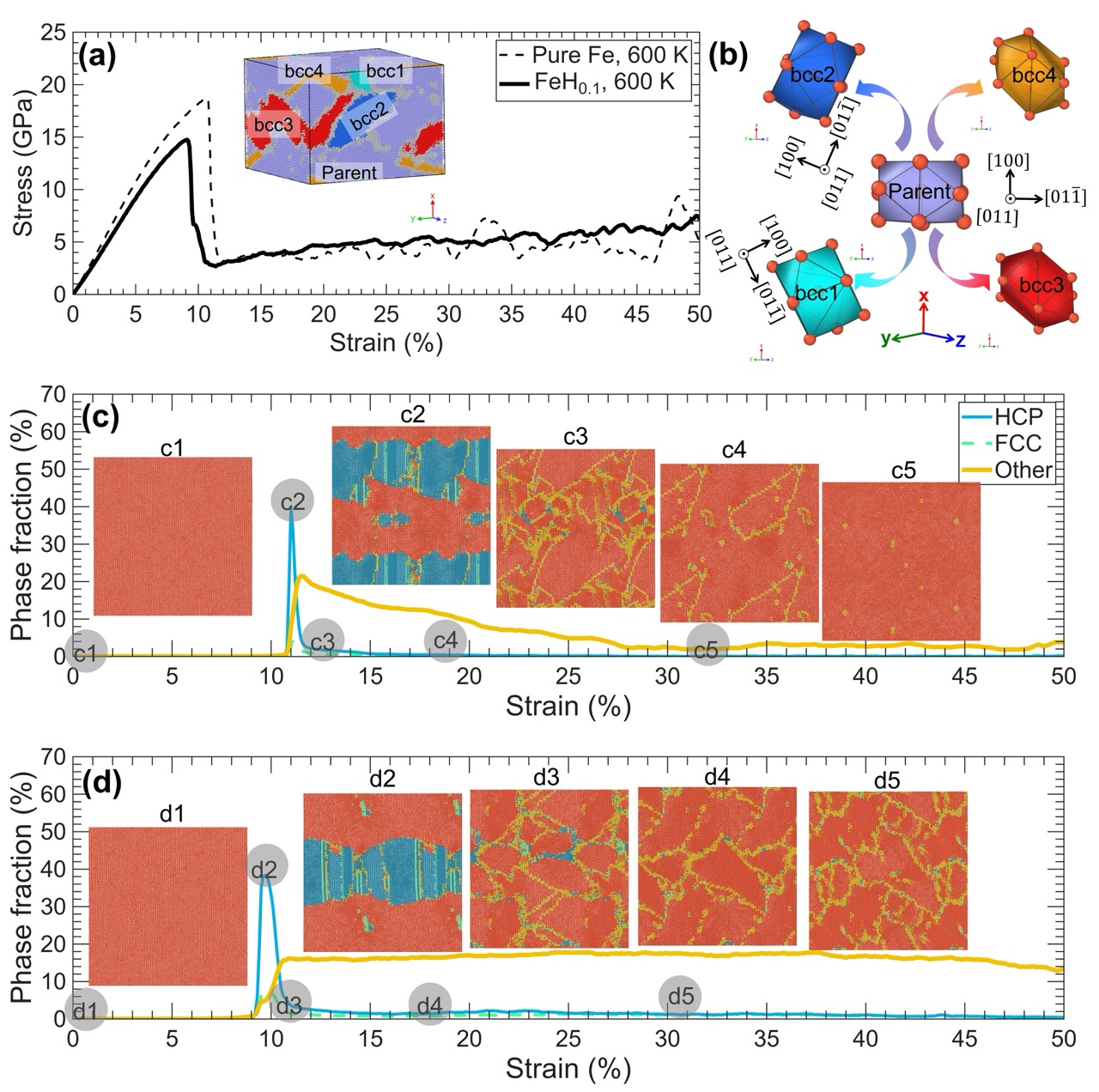}
    \caption{
    Compressive deformation and phase evolution of pure Fe and $\mathrm{FeH}_{0.1}$ under $[100]$ compression at 600~K and a strain rate of $10^{9}$~s$^{-1}$.
    (a) Stress--strain curves and a representative three-dimensional view of the transformation-generated microstructure.
    (b) Crystallographic relationships between the parent bcc lattice and the four bcc twin variants.
    (c,d) Evolution of the hcp, fcc, and structurally disordered (\textit{Other}) phase fractions in pure Fe and $\mathrm{FeH}_{0.1}$, respectively. The bcc fraction, which constitutes the remainder of the system, is omitted to better resolve the minority phases. The snapshots show representative microstructures at the strains marked on the phase-fraction curves and are viewed along $[0\bar{1}1]$.
    }
    \label{fig:compression_600}
\end{figure}

Figure~\ref{fig:compression_600} shows the mechanical response and microstructural evolution of pure Fe and $\mathrm{FeH}_{0.1}$ under $[100]$ compression at 600~K. In both systems, deformation proceeds through transformation-mediated twinning via the bcc$\rightarrow$hcp$\rightarrow$bcc pathway. Near the peak stress, a large fraction of the bcc lattice abruptly transforms into the hcp phase. The hcp fraction then decreases rapidly as the intermediate phase transforms back into multiple $\{112\}$ twin variants. Compared with the corresponding results at 300~K in Fig.~\ref{fig:stress_strain}, both pure Fe and $\mathrm{FeH}_{0.1}$ exhibit lower peak stresses and earlier phase transformation at 600 K. Formation of the hcp intermediate requires correlated atomic shuffling and local shear within the highly stressed bcc lattice. At 600~K, thermal fluctuations enable the system to sample transformation-compatible configurations more readily and reduce the mechanical work required to overcome the nucleation barrier. The transformation therefore begins at a lower applied stress and strain at 600 K than at 300~K.

As shown in Fig.~\ref{fig:compression_600}c, multiple $\{112\}$ twin variants in pure Fe subsequently detwin, similar to the case at 300 K. However, the elevated temperature alters the effect of hydrogen on the resulting twin network and flow stress significantly. At 300~K, hydrogen markedly prolongs the hcp-mediated transformation and promotes the growth and impingement of all four $\{112\}$ twin variants, leading to a highly organized network of mutually constraining  $\{332\}$ and $\{10\,9\,3\}$ twin boundaries. In contrast, at 600~K, the hcp fraction is consumed over a relatively narrow strain interval. As shown by configurations d3--d5 in Fig.~\ref{fig:compression_600}d, the resulting microstructure is less interconnected than the multimodal  twin network observed at 300~K. Thus, elevated temperature limits the development and stabilization of the high-index twin network. Consequently, the flow stress at 600 K no longer exhibits the pronounced elevation observed at 300~K. 

This observation suggests a competition between hydrogen and temperature in governing the twinning behavior. On one hand, increased thermal fluctuations facilitate dislocation-mediated processes, thereby reducing the propensity for twinning. On the other hand, the transient hcp intermediate is a low-temperature, high-pressure phase of Fe, such that its stability decreases with increasing temperature. Consequently, at 600 K, elevated temperature suppresses twin-variant coarsening and inhibits the formation of a stable multimodal  twin network. In contrast, at 300 K, hydrogen promotes the development of a persistent multimodal  twin network that contributes substantially to the high flow stress.

\subsection{Effect of loading direction}
\label{sec:tension_600}

\begin{figure}[H]
    \centering
    \includegraphics[width=0.95\textwidth]{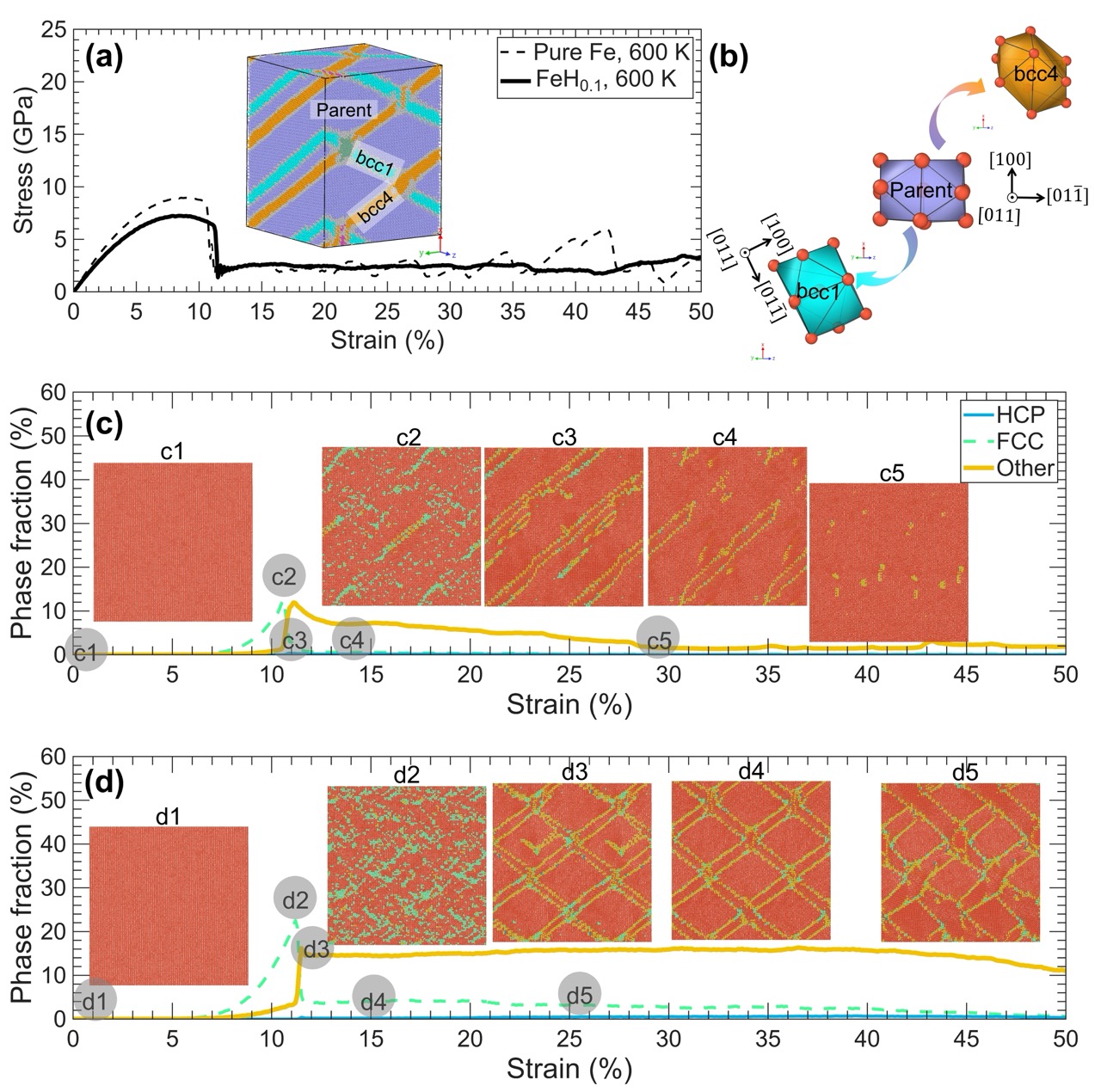}
    \caption{
    Tensile deformation and phase evolution of pure Fe and $\mathrm{FeH}_{0.1}$ under $[100]$ tension at 600~K and a strain rate of $10^{9}$~s$^{-1}$.
    (a) Stress--strain curves and a representative three-dimensional view of the transformation-generated twin structure. (b) Crystallographic relationships between the parent bcc lattice and the two bcc twin variants. (c,d) Evolution of the hcp, fcc, and structurally disordered (\textit{Other}) phase fractions in pure Fe and $\mathrm{FeH}_{0.1}$, respectively. The bcc fraction is omitted for clarity. The snapshots show representative microstructures at the strains marked on the phase-fraction curves and are viewed along $[0\bar{1}1]$.
    }
    \label{fig:tension_600K}
\end{figure}

Based on our previous work, reverse loading in bcc materials is known to produce pronounced tension–compression asymmetry, distinct intermediate phases, and different $\{112\}$ twinning modes~\cite{zahiri2024anisotropy}. In this section, we examine the deformation under tensile loading. Figure~\ref{fig:tension_600K} shows the stress--strain response and phase evolution of pure Fe and $\mathrm{FeH}_{0.1}$ subjected to $[100]$ tension at 600~K. Near the peak stress, fcc-like domains nucleate within the strained bcc lattice. Afterwards, the reverse transformation initially generates two $\{112\}$ twin variants, as illustrated by configurations c2 and c3 in Fig.~\ref{fig:tension_600K}c. The twin microstructure subsequently coarsens and then undergoes detwinning, as shown in c4 and c5. The gradual decrease in the \textit{Other} fraction reflects the accompanying reduction in the density of twin boundaries.

A markedly different evolution occurs in $\mathrm{FeH}_{0.1}$. As shown in configurations d2--d5 of Fig.~\ref{fig:tension_600K}d, the transformation-generated variants undergo extensive growth and intersect, forming a cross-linked twin-boundary network between bcc1 and bcc4. Similar to the compression case, hydrogen stabilizes the $\{112\}$ twin variants, enabling them to persist and grow over substantial distances. The persistent \textit{Other} fraction provides quantitative evidence for the large population of atoms associated with twin boundaries, boundary steps, and twin--twin junctions. 

\begin{figure}[H]
    \centering
    \includegraphics[width=0.8\textwidth]{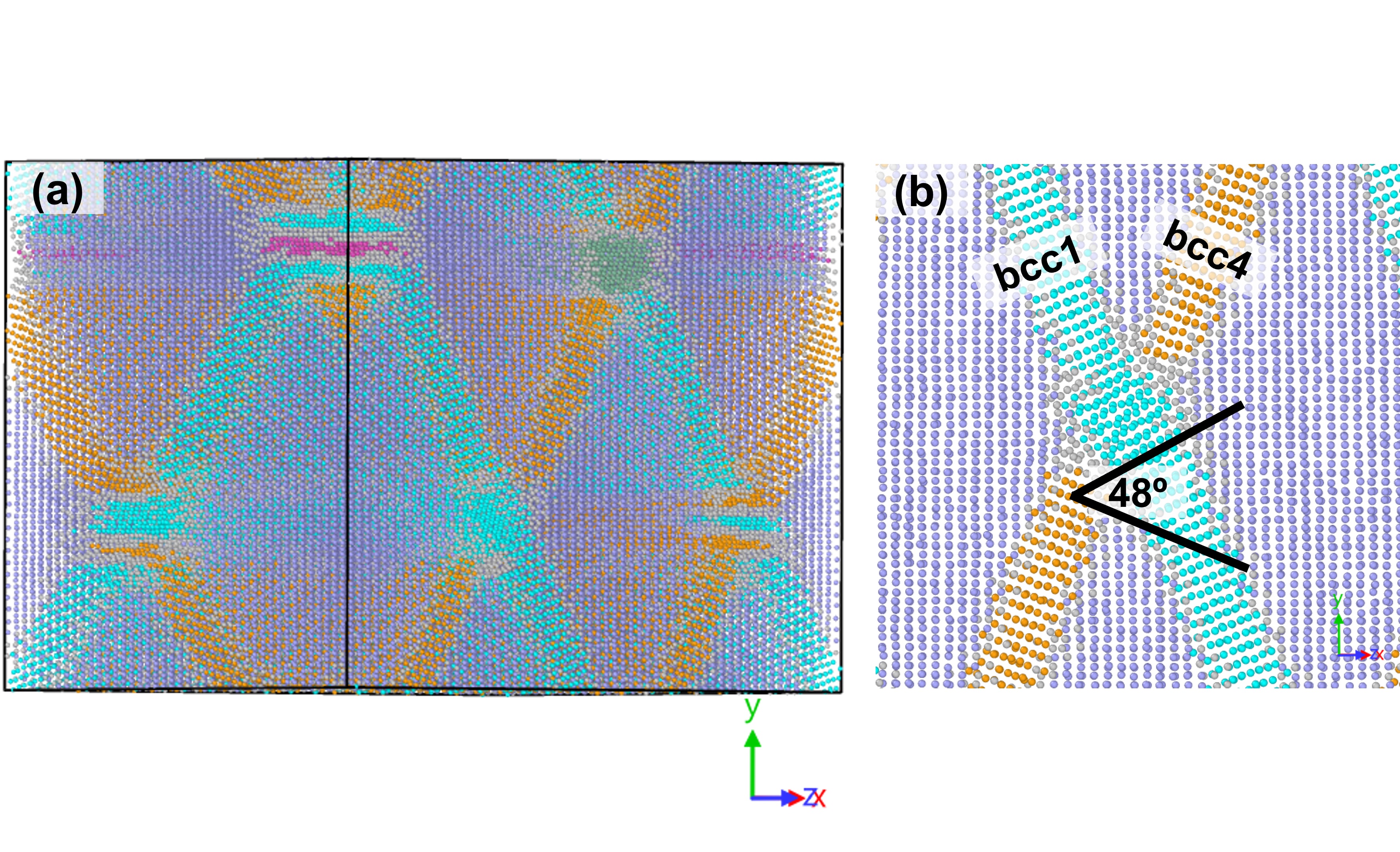}
    \caption{(a) Grain-segmentation analysis of H-containing Fe during $[100]$ tension at 600~K. (b) bcc1 and bcc4 form an approximately $48^{\circ}$ misorientation. }
    \label{fig:sg_fcc}
\end{figure}

In both tension and compression, the hydrogen-assisted stabilization of the $\{112\}$ twin variants promotes their growth, thereby increasing the likelihood of variant impingement and twin--twin interactions. However, $\{112\}$ twin variants formed through different intermediate phases, and thus via distinct transformation pathways, may exhibit different twinning elements. Consequently, the nature of twin--twin interactions may differ between tension and compression. Figure~\ref{fig:sg_fcc} shows two surviving bcc variants in $\mathrm{FeH}_{0.1}$ at 600~K with a misorientation of approximately $48^{\circ}$. Both the misorientation angle and our theoretical calculations indicate that the non-co-zone variant formed under tension corresponds to a $\{7\,4\,1\}$ twin. To the best of our knowledge, this high-index twin has previously been predicted only in the theoretical calculations of Gao et al.~\cite{gao2020twinningpath}, whereas the present study provides the first direct observation of its dynamic formation process.

For the theoretical calculation, we monitor the evolution of three orthogonal directions throughout the twin formation process. This tracking leads us to derive a specific set of orientation relations for bcc1, given in Eq.~\eqref{eq:or1}.

\begin{equation}
\begin{split}
    &[100]_{bcc} \quad\leftrightarrow\quad [100]_{fcc} \quad\leftrightarrow\quad [011]_{bcc1}
    \\[2mm]
    &[011]_{bcc} \quad\leftrightarrow\quad [010]_{fcc} \quad\leftrightarrow\quad [\bar{1}00]_{bcc1}
    \\[2mm]
    &[0\bar{1}1]_{bcc} \quad\leftrightarrow\quad [001]_{fcc} \quad\leftrightarrow\quad [0\bar{1}1]_{bcc1}\,.
\end{split}
\label{eq:or1}
\end{equation}

Accordingly, we define the parent basis as $\mathbf{G}_1 = a_0\,\mathbf{e}_1$, $\mathbf{G}_2 = a_0\,(\mathbf{e}_2+\mathbf{e}_3)$, and $\mathbf{G}_3 = a_0\,(-\mathbf{e}_2+\mathbf{e}_3)$, and the twin basis as $\mathbf{g}_1 = a_0\,(\mathbf{e}_2+\mathbf{e}_3)$, $\mathbf{g}_2 = -a_0\,\mathbf{e}_1$, and $\mathbf{g}_3 = a_0\,(-\mathbf{e}_2+\mathbf{e}_3)$. Hence, the correspondence matrix $C_{ij}$ for this deformation is
\begin{equation}
\begin{split}
C_{ij}|_{bcc-fcc-bcc1} = \left[
      \begin{array}{ccc}
        0 \quad & \quad -0.5 \quad & \quad -0.5\\
        1 \quad & \quad 0.5 \quad & \quad -0.5\\
        1 \quad & \quad -0.5 \quad & \quad 0.5\\
      \end{array}
    \right]\,.
\end{split}
\label{eq:dg}
\end{equation}

In addition, based on the lattice correspondence for bcc4, we identify the following correspondence matrix

\begin{equation}
\begin{split}
C_{ij}|_{bcc-fcc-bcc4} = \left[
      \begin{array}{ccc}
        0.5 \quad & \quad -1 \quad & \quad 0.5\\
        0.5 \quad & \quad 0 \quad & \quad -0.5\\
        0.5 \quad & \quad 1 \quad & \quad 0.5\\
      \end{array}
    \right]\,.
\end{split}
\label{eq:cfbcc3}
\end{equation}

For twin--twin interactions between bcc1 and bcc4, the stretch $\mathbf{U}_I$ is calculated from the total correspondence $\mathbf{C}_{bcc-hcp-bcc1}\cdot\mathbf{C}_{bcc-hcp-bcc4}^{-1}$. Based on this stretch, from Eq.~\eqref{eq:te3} we obtain
\begin{equation}
\begin{split}
    \mathbf{K}_1 = (\bar{4}7\bar{1})\,,\quad
    \boldsymbol\eta_1 = [113]\,,\quad
    s_1 =1.2247\,,
    \\[2mm]
    \mathbf{K}_2 = (0\bar{1}\bar{1})\,,\quad
    \boldsymbol\eta_2 = [1\bar{1}1]\,,\quad
    s_2 =1.2247\,.
    \label{eq8}
\end{split}
\end{equation}
Indeed, the $\{7\,4\,1\}$ twin is characterized by a misorientation of $48^{\circ}$ about $\langle1\bar{1}0\rangle$, consistent with the observation in the twin--twin interaction in Fig.~\ref{fig:sg_fcc}a~\cite{gao2020twinningpath}.

\section{Discussion}
\label{sec:discussion}

Under compression, the deformation proceeds through transformation-mediated twinning via the bcc$\rightarrow$hcp$\rightarrow$bcc pathway, whereas under tension, it proceeds through the bcc$\rightarrow$fcc$\rightarrow$bcc pathway. Different intermediate phases indicate different deformation pathways and, consequently, different twin--twin interactions. In the hcp-mediated case, a $\{10\,9\,3\}$ high-index twin is observed between bcc1/bcc4 or bcc2/bcc3, whereas in the fcc-mediated case, a $\{7\,4\,1\}$ high-index twin is observed between bcc1/bcc3. Although both arise from interactions between non-co-zone $\{112\}$ twin variants, these high-index twin--twin boundaries are significantly different. This difference originates from the fact that despite sharing the same habit plane, the $\{112\}$ twins formed over different intermediate phases differ in the shear direction, shear magnitude, and shuffle content. 

As described in detail in our previous work~\cite{zahiri2024anisotropy}, the bcc1 twin variant generated through the fcc intermediate under tension is characterized by $\mathbf{K}_1=(211)$, $\boldsymbol{\eta}_1=[1\bar{1}\bar{1}]$, and a twinning shear of $s=\sqrt{2}/2=0.7071$. In contrast, the bcc1 twin variant generated through the hcp intermediate under compression follows $\mathbf{K}_1=(211)$, $\boldsymbol{\eta}_1=[\bar{1}11]$, and $s=\sqrt{2}/4=0.3536$. Thus, relative to the fcc-mediated mode, the hcp-mediated mode reverses the shear direction and reduces its magnitude by a factor of two. The two twins should therefore not be regarded as mechanically equivalent realizations of a generic $\{112\}$ twin. Their final orientation relationships may be closely related, but their atomic displacement fields and their interactions with neighboring variants are fundamentally different.

\begin{figure*}[!htbp]
    \centering
    \includegraphics[width=\textwidth]{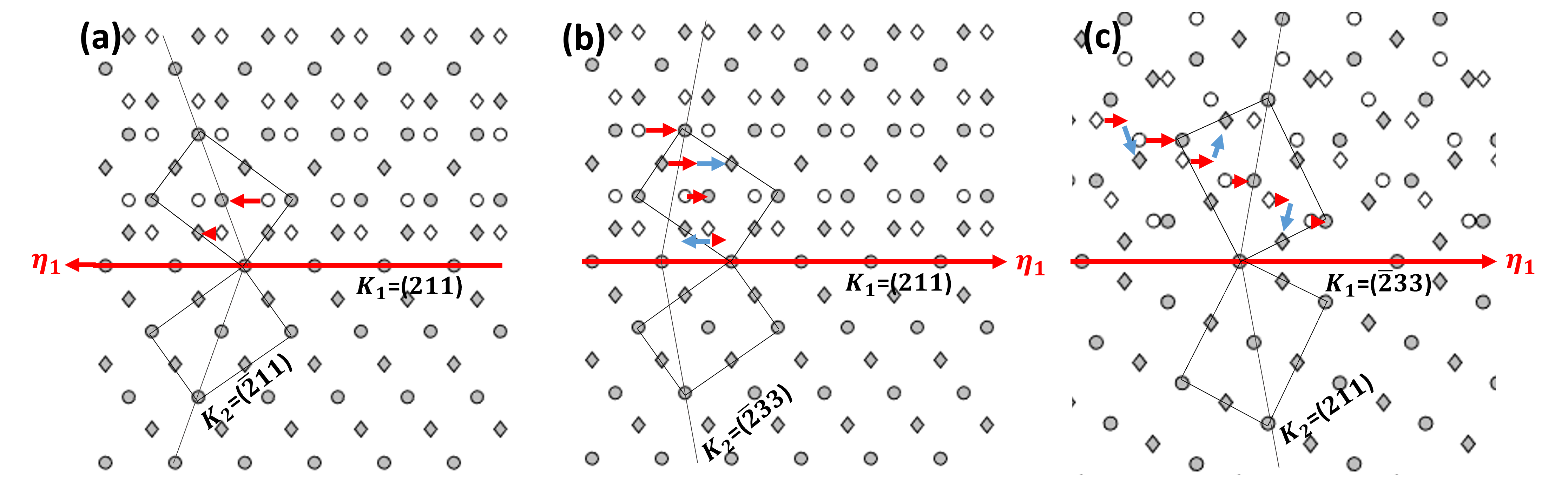}
    \caption{
    Dichromatic complexes of the transformation-generated twin modes projected along $\langle1\bar{1}0\rangle$:
    (a) the no-shuffle $\{112\}$ twin generated through the fcc-mediated pathway;
    (b) the 1/2 atoms shuffle $\{112\}$ twin generated through the hcp-mediated pathway; and
    (c) its reciprocal $\{332\}$ twin mode.
    The lower parent bcc lattice is represented by gray symbols, the superposed parent lattice by open symbols, and the twin lattice by the upper gray symbols. Red arrows indicate the homogeneous shear along $\boldsymbol{\eta}_1$, while blue arrows denote the additional atomic shuffles required to complete the lattice correspondence. Adapted from~\cite{zahiri2024anisotropy} with permission from Elsevier.
    }
    \label{fig:di}
\end{figure*}

The atomic origin of this distinction becomes evident from the dichromatic complexes in Fig.~\ref{fig:di}. In the fcc-mediated tensile mode shown in Fig.~\ref{fig:di}a, the parent lattice can be mapped onto the twin lattice by a homogeneous shear without an additional sublattice shuffle. All atoms are displaced directly toward crystallographically equivalent sites through the shear associated with $\boldsymbol{\eta}_1$. This is the conventional no-shuffle $\{112\}\langle111\rangle$ twinning mode widely discussed for bcc Fe, molybdenum, tungsten, and related metals~\cite{ChristianMahajan1995,gao2020twinningpath}. The absence of shuffle is consistent with the nature of the bcc--fcc correspondence. The Bain transformation is a homogeneous lattice distortion and does not require alternating atomic planes to undergo distinct internal displacements. When the transient fcc structure transforms back into bcc, the resulting parent--product relationship can therefore be represented by a homogeneous twinning shear with
$s=\sqrt{2}/2$.

The hcp-mediated compression mode is fundamentally different. As shown in Fig.~\ref{fig:di}b, homogeneous shear alone does not map every parent atom onto a correct twin-lattice site. One-half of the atoms require an additional shuffle relative to the sheared configuration. This atomic displacement pattern follows directly from the bcc--hcp Burgers pathway, which combines a lattice shear with relative shuffling of alternating atomic planes. The reverse hcp$\rightarrow$bcc transformation consequently produces a 1/2 atoms shuffle $\{112\}$ twin with a shear magnitude of only $\sqrt{2}/4$. Part of the required parent--product correspondence is supplied by the discrete shuffle, reducing the homogeneous component of the twinning deformation. The 1/2 atoms shuffle $\{112\}$ mode is crystallographically reciprocal to the $\{332\}$ mode shown in Fig.~\ref{fig:di}c, which has been reported extensively in metastable Ti alloys~\cite{xiao2022effect,zhou2017accommodative,kvashin2021migration}. 

These findings show that the final twin boundary alone does not uniquely reveal the mechanism by which it formed. Twins with nominally similar $\{112\}$ orientation relationships can possess different shear magnitudes, shuffle contents, and interaction behaviors because they descend from different intermediate structures. Resolving the transient transformation pathway is therefore essential for identifying the operative twinning mechanism and predicting the resulting variant hierarchy. This interpretation is consistent with growing evidence that transitional structures and symmetry-breaking transformation pathways play a decisive role in twinning and variant selection~\cite{gao2020twinningpath,li2023transitional,zahiri2024anisotropy}.

As summarized in Table~\ref{table:all}, the loading direction selects the intermediate phase and the transformation pathway. Under compression, the hcp intermediate generates 1/2 atoms shuffle $\{112\}$ parent--product twins. Co-zone twin--twin interactions form $\{332\}$ boundaries, whereas non-co-zone twin--twin interactions form $\{10\,9\,3\}$ boundaries. Under tension, the fcc intermediate produces no-shuffle $\{112\}$ twins. Co-zone variants can merge completely~\cite{zahiri2024anisotropy}, while non-co-zone variants form $\{7\,4\,1\}$ boundaries. The different high-index boundaries are therefore direct crystallographic records of the different intermediate phases.

\begin{table}[htbp]
    \centering
    \caption{Transformation pathways, twinning modes, and twin--twin interactions observed in the DNN-MD simulations.}
    \vspace{2mm}
    \label{table:all}
    \begin{tabular}{|>{\raggedright\arraybackslash}p{0.28\textwidth}|>{\raggedright\arraybackslash}p{0.28\textwidth}|>{\raggedright\arraybackslash}p{0.28\textwidth}|}
        \hline
        \textbf{Characteristic}
        & \textbf{$[100]$ compression}
        & \textbf{$[100]$ tension} \\
        \hline
        Transformation pathway
        & bcc$\rightarrow$hcp$\rightarrow$bcc
        & bcc$\rightarrow$fcc$\rightarrow$bcc \\
        \hline
        $\{112\}$ twin mode
        & 1/2 atoms shuffle $\{112\}$ twin & No-shuffle $\{112\}$ twin \\
        \hline
        Twinning elements
        & $\mathbf{K}_1=(211)$, $\boldsymbol{\eta}_1=[\bar{1}11]$
        & $\mathbf{K}_1=(211)$, $\boldsymbol{\eta}_1=[1\bar{1}\bar{1}]$ \\
        \hline
        Twinning shear
        & $s=\sqrt{2}/4=0.3536$
        & $s=\sqrt{2}/2=0.7071$ \\
        \hline
        Co-zone twin--twin interaction
        & $\{332\}$ boundary
        & merging of impinging variants \\
        \hline
        Non-co-zone twin--twin interaction
        & $\{10\,9\,3\}$ boundary
        & $\{7\,4\,1\}$ boundary \\
        \hline
    \end{tabular}
\end{table}

This framework distinguishes the crystallographic availability of a boundary from its \emph{kinetic realization}. The intermediate phase determines the set of product variants and the possible interfaces among them. Hydrogen and temperature determine whether those variants survive long enough to grow and interact. Thus, the transformation pathway defines the available crystallographic hierarchy, whereas the deformation conditions determine which part of that hierarchy becomes visible in the final microstructure.

This distinction explains why the same underlying transformation pathway can produce markedly different microstructures at different temperatures or hydrogen concentrations. Rapid elimination of unfavorable variants leaves only isolated $\{112\}$ twins or a single dominant product. Prolonged coexistence permits co-zone and non-co-zone variants to impinge, producing $\{332\}$, $\{10\,9\,3\}$, or $\{7\,4\,1\}$ interfaces. High-index twinning therefore does not constitute an independent nucleation event. It emerges only after the primary transformation-generated variants have developed sufficiently to interact.

\section{Conclusions}
\label{sec:conclusions}
A high-accuracy DNN potential for Fe-H was trained using ab initio molecular dynamics data. Using DNN-MD simulations, we investigated the effects of hydrogen, loading direction, and temperature on the deformation behavior of bcc Fe. Deformation leads to the extensive formation of multiple $\{112\}$ twin variants through a transient intermediate phase. Hydrogen reduces the stress required for the phase transformation by promoting local structural transformations; however, its dominant post-yield effect is to modify twin--twin interactions and the stability of twin variants. Under compression at 300~K, hydrogen stabilizes multiple bcc variants and facilitates the formation of a persistent, mutually pinning high-index $\{332\}$ and $\{10\,9\,3\}$ twin-boundary network, thereby suppressing detwinning and substantially increasing the flow stress. In addition, elevated temperature lowers the transformation stress, but its effect on the final twin structure is even more pronounced. At 600~K, thermal activation partially counteracts the effect of hydrogen, allowing twin interfaces to migrate and reorganize more readily. Consequently, the stable multimodal  twin network does not form, and no corresponding increase in flow stress is observed. These results highlight the critical role of a stable multimodal  twinning network in sustaining high flow stress.

Moreover, we revealed distinct tension--compression asymmetry in Fe and $\mathrm{FeH}_{0.1}$. Under $[100]$ compression, twinning proceeds through a bcc$\rightarrow$hcp$\rightarrow$bcc pathway, whereas tension activates a bcc$\rightarrow$fcc$\rightarrow$bcc pathway. These distinct intermediates and transformation pathways generate different $\{112\}$ twinning modes, with different shear magnitudes and shuffle characteristics, and consequently produce different twin--twin interactions: $\{332\}$ and $\{10\,9\,3\}$ boundaries under compression, and $\{7\,4\,1\}$ boundaries under tension. Notably, this work provides the first direct observation of the dynamic formation of these high-index twin boundaries in MD simulations. Furthermore, by combining the atomistic observations with theoretical calculations, we determine the complete twinning elements of these previously less-known twinning modes.

Overall, these results establish phase transformation as the underlying mechanism that governs twin selection in dynamically loaded Fe. The transient intermediate phase dictates the accessible bcc variants and their crystallographic relationships, whereas hydrogen and temperature determine how these variants grow, interact, detwin, and ultimately evolve into stable high-index twin networks. This framework provides a mechanistic basis for understanding hydrogen-induced hardening, tension–compression asymmetry, and the temperature dependence of twinning in bcc Fe, which should also apply readily to other bcc materials.

\section*{Acknowledgments}
This work was supported by the National Science Foundation (NSF) under Grant No.\ DMR-2240125. The authors would also like to acknowledge the support of Research \& Innovation and the Office of Information Technology at the University of Nevada, Reno, for computing time on the Pronghorn High-Performance Computing Cluster.

\section*{Data Availability Statements}
The data that support the findings of this study are available from the corresponding author upon reasonable request. 

\bibliographystyle{elsarticle-num}
\bibliography{references}

\end{document}